\documentstyle[12pt]{article}
\begin{document}
\newcommand{\vsone}{\vspace{1cm}}
\newcommand{\be}{\begin{equation}}
\newcommand{\ee}{\end{equation}}
\newcommand{\bea}{\begin{eqnarray}}
\newcommand{\eea}{\end{eqnarray}}
\newcommand{\pr}{\paragraph{}}

\newcommand{\nd}[1]{/\hspace{-0.6em} #1}
\begin{titlepage}
\begin{flushright}
CERN-TH.6413/92 \\
ACT-2/92 \\
CTP-TAMU-12/92\\
\end{flushright}
\begin{centering}
\vspace{.1in}
{\large {\bf Measuring the $W$-hair  of String
Black Holes }} \\

\vspace{.4in}
{\bf John Ellis} and {\bf N.E. Mavromatos}\\
\vspace{.05in}
Theory Division, CERN, CH-1211, Geneva 23, Switzerland  \\
and \\

\vspace{.05in}
{\bf D.V. Nanopoulos}\\

\vspace{.05in}
Center for Theoretical Physics, Dept. of Physics, \\
Texas A \& M University, College Station, TX 77843-4242, USA \\
\vspace{.03in}
and \\
\vspace{.05in}
Astroparticle Physics Group \\
Houston Advanced Research Center (HARC),\\
The Woodlands, TX 77381, USA\\
\vspace{.1in}
{\bf Abstract} \\
\vspace{.05in}
\end{centering}
{\small
\paragraph{}
We have argued previously that the infinitely many
gauge symmetries of string theory provide an infinite set
of conserved (gauge)
quantum numbers ($W$-hair) which characterise
black hole states and maintain quantum coherence. Here
we
study ways of measuring
the
$W$-hair of spherically-symmetric four-dimensional objects
with event horizons, treated as effectively
two-dimensional
string black holes.
Measurements can be done either through the
s-wave
scattering of
light particles
off the string black-hole background,
or
through interference
experiments of
Aharonov-Bohm
type.
In the first type of measurement,
selection rules
restrict
the number of
particles
emitted by the black hole,
while in the second method  the wave-functions of
fundamental
strings
scattered off the black hole have
phases
that
can be
measured in appropriate
interference experiments.}
\par
\vspace{0.2in}
\begin{flushleft}
CERN-TH.6413/92 \\
ACT-2/92 \\
CTP-TAMU-12/92 \\
February 1992 \\
\end{flushleft}
\end{titlepage}
\newpage
\section{Introduction}
\paragraph{}
The reconciliation of general relativity with
quantum mechanics is one of the key problems
in physics. One of its many facets
is the quantisation of gravitational
effects in a flat space-time background, including
the calculability of perturbation theory, which should be
{\it finite} or at least {\it renormalisable}. Another
facet is the problem of quantisation in non-flat
space-times that are solutions of the gravitational
field equations. Still another is the full non-perturbative
treatment of fluctuations in space-time. And so on...
\pr
Conventional point-like quantum field theory has been
unable to resolve any of these problems, whilst string
theory is an ambitious candidate for a framework in which
they can all be resolved. Indeed, it has now been established
that string perturbation theory in a flat space-time background
is {\it finite} \cite{mand}. On the other hand, the full
non-perturbative treatment of fluctuations in space-time
requires the development of a fully-fledged string field
theory, which so far only exists for simplified toy models
of string gravity coupled to matter \cite{das,sen}.
In a series of recent papers \cite{emn1,emn2,emn3,emn4},
we have been studying the intermediate problem of quantum
theory in a non-flat space-time background.
\pr
Specifically, we have been investigating whether
conventional quantum coherence can be maintained in a black hole
background in string theory. Semiclassical arguments \cite{hawk}
in conventional
point-like quantum field theory indicate that macroscopic
black holes behave like mixed, thermal states, which has
motivated suggestions \cite{hawpop} that quantum coherence
{\it cannot} be maintained at a fundamental level
when microscopic non-perturbative
fluctuations in space-time are taken into account. The root
of this problem with quantum coherence is the observation
that the apparent entropy $S$ of a black hole is proportional
to the area of its event horizon \cite{hawk,bek}: $S \propto  A$,
which is in turn $\propto  M^2$ for an uncharged axisymmetric
black hole. Thus the apparent entropy of a black hole is
unbounded, whereas any point-like quantum field theory
has only a finite set
of quantum numbers. These are insufficient
to encode all the information carried by matter
that collapses gravitationally,
and thereby distinguish
all the states of a black hole, which must therefore be described
by a {\it mixed} state.
\pr
However, we have argued \cite{emn1,emn2}
that there are infinitely many {\it gauge} symmetries
in string theory, which lead to an {\it infinity} of
{\it conserved} quantities, ``$W$-hair'', that maintain
quantum coherence in the presence of a black hole.
First demonstrated in a two-dimensional model, this
argument applies also to four-dimensional black holes, as
has been shown explicitly in the spherically-symmetric
case of physical interest \cite{emn4}. It has been shown in
general \cite{emn3}
that,
within the context of string theory,
black hole decay can be calculated as a conventional
quantum-mechanical process which does not involve thermal
or other mixed states, and the number of pure solitonic states
of a stringy black hole has been estimated \cite{emn4} and shown
to correspond to the number of black hole states found previously
by semi-classical field-theoretic arguments \cite{hawk,bek}. Thus
the entropy
of a four-dimensional black hole is only {\it apparent}, and can in
principle be reduced to zero by measurements of the intrinsically
stringy $W$-hair quantum numbers of these solitonic states.
\pr
The question then arises, how could one imagine
measuring these quantum numbers in practice?
In this paper we propose two types of possible measurement.
One is via selection rules for the scattering of light particles
off stringy black holes, and the other is an infinite set of
stringy Aharonov-Bohm effects.
\pr
The concept of the first type of measurement
is very similar to that of $\pi$-nucleon scattering in the
Skyrme model \cite{skyrm}. In that case the $\Delta(1232)$ $3-3$
resonance, for example, is a higher spin (and isospin) soliton
which can be excited if the $\pi$ energy is resonant, and its
decay satisfies certain selection rules. This picture has been
extended
to higher resonances, with soliton calculations
reproducing well the phase shifts in different partial
waves \cite{km} and general selection
rules derived \cite{mb}. In our case, there is an infinite
set
of black hole soliton states, classified by the quadratic
Casimir and `magnetic' quantum numbers of an internal symmetry
group, which are excited at calculable energies and decay into
distinctive numbers of light final-state particles. These results
are derived in the limiting case of a flat space-time background
that represents
the end-point of black hole decay, and then
in the generic (potentially macroscopic) black hole case.
The key r\^ ole of the $s$-wave dynamics and the particle
production selection rules are reminiscent of the
Callan-Rubakov \cite{caru} process in scattering off a monopole.
\pr
The second type of measurement involves the characteristic
inteferences between states propagating in the neighbourhood of,
and far from,
a black hole. Since there are infinitely
many massive string states which can be scattered in this way, there
is an infinity of such possible measurements, albeit
with certain practical difficulties in the case of a macroscopic
black hole.
\pr
The layout of this paper is as follows. In section 2 we review
the infinite set of stringy $W$-symmetries on the world-sheet
and their elevation to physical space-time symmetries. In section 3
we derive the selection rules for scattering in a flat space-time
background \cite{klepol}, and extend these to the generic
black hole case in section 4. Section 5 presents the generalised
Aharonov-Bohm measurements, and we discuss some outstanding
issues in section 6.
\pr
\section{W-symmetries on the world-sheet
and in physical space-times}
\paragraph{}
We begin our study with
a brief review of
the $W$-symmetries possessed by
string theories
in two-dimensional
space-time, and hence
also
string theories in spherically-symmetric
four-dimensional space-time.
\pr
The first notion of a target space-time $W$-symmetry
was presented by Avan and Jevicki \cite{ava}
in the context of the collective field representation
of the $c=1$ matrix model \cite{das}. The
latter is nothing other than
string field theory for the only propagating degree of freedom
of the two-dimensional strings, the so-called
`tachyon', which is actually massless in such theories.
The existence of an infinite-dimensional Cartan subalgebra
of conserved charges was
demonstrated.
Subsequently,
it was
suggested \cite{emn1} that these symmetries could be understood
as `hidden' gauge
symmetries \cite{ven}
of the underlying
string theory,
associated with the higher excited string states, which in two dimensions
are non-propagating and have definite values of energy and momentum.
These states had been known to exist in intermediate channels
of tachyon scattering amplitudes in matrix models \cite{gross}, in
Das-Jevicki theory \cite{dem},
and in
$c=1$
Liouville
theory
\cite{pol}, but their physical
significance had not been realised at the time.
It is by now clear that these modes
are essential for the perturbative
unitarity of flat-space string scattering matrix  \cite{sak}.
In fact, it is through the usual factorisation via the
operator product expansion (OPE)
of vertex operators
that these discrete states are produced as intermediate
states in tachyon
scattering amplitudes.
Soon after the suggestion of ref. \cite{emn1}
on the association of $W$-symmetries with higher-level string states,
Moore and Seiberg \cite{ms}
constructed explicitly a $W_{1+\infty}$ algebra of
symmetries in the fermionic representation of $c=1$ matrix models,
and showed its connection
with the higher-spin (discrete)
string modes. In view of the target-space interpretation
of the model
as a field theory of the tachyon field \cite{das}, these
charges characterise
the flat string background in two dimensions, or the $s$-wave
sector of
dimensionally-reduced four-dimensional strings \cite{emn4}.
\paragraph{}
Although the
target-space interpretation of these symmetries
was evident
by construction,
it is
still
useful
to
understand their origin as
world-sheet symmetries, which would then be elevated to
target-space ones in the usual way, via induced {\it canonical}
deformations
of the corresponding conformal field theory \cite{ovr}.
\paragraph{}
Let us be more precise. String theory in first-quantised form
is formulated as a theory on the world-sheet.
Hence, at a
superficial
level the only apparent symmetries
are those on the world-sheet.
One is
then
interested in knowing under which conditions
such symmetries
could be
elevated into symmetries of the physical
space-time.
Let
$h=\int d\sigma j(\sigma)$ be a conserved charge of
a current $j(\sigma)$ generating such
a symmetry.
This implies
the invariance of the
Fradkin-Tseytlin
generating function for amplitudes under
an appropriate change of the $\sigma$-model fields (target-space
coordinates) that corresponds to the symmetry in question \cite {ven}.
The parameters of the transformation may, and in fact
do in
our case \cite{emn2}, have an explicit dependence on world-sheet
coordinates. This makes this formalism unwieldy for
getting non-perturbative results in some
closed form. As an alternative, Evans and Ovrut \cite{ovr}
suggested the study of hidden symmetries through the induced
deformations on the stress-tensor of the conformal field theory
in a certain background \{ g \}. If $T_g$
(${\overline T_g} $) denotes
the holomorphic (antiholomorphic) part of the
stress
tensor of the $\sigma$-model, then the induced (infinitesimal)
deformation is $ \delta T_g = i[h, T_g]$. The deformation is
a symmetry of the physical (target)
space-time if

\be
    \delta T_g = T_{g+\delta g} - T_g
\label{symm}
\ee

\noindent for some induced transformation of the couplings:
$g$  $ \rightarrow $  $g + \delta g $. A
deformation
is
said to be
{\it canonical} if \cite{ovr}

\be
     \delta T_g = \Phi _{(1,1)}
\label{can}
\ee

\noindent where $\Phi _{(1,1)}$ is a primary field of dimension
$(1,1)$. Thus conformal invariance is automatically satisfied
for canonical deformations. Due to the completeness of the
set of $(1,1)$ vertex operators in string theory, it is also
evident, in view of (\ref{symm}), that a canonical deformation
is also a {\it symmetry} of target space-time.
An important comment is that, if the current generator of the
world-sheet symmetry is an operator of conformal dimension
$(1,0)$ or $(0,1)$, then the induced deformation has conformal
dimension $(1,1)$ and, hence, generates a target space
symmetry of the background. This is precisely what happens in the
case of two-dimensional string theory in flat space-times
\cite{witt2}. There is an infinity of $(1,0)$ or
$(0,1)$
world-sheet currents,
constructed out of states of non-standard ghost number,
which generate $W$-symmetries on the world-sheet
that can be lifted \cite{emn2}
to $W$-symmetries of the
physical space-time
of two-dimensional string theory, formulated on
flat
backgrounds. These symmetries have the properties that
they leave invariant
under
target
time evolution
the two-dimensional
tachyon
phase-space,
which, in view of \cite{das},
is
equivalent to a matrix-model phase space \cite{witt2}.
Following
indications about
background independence of the Das-Jevicki string feld theory
\cite{das,alw,emn2},
one might expect \cite{emn2}
that
these $W$-symmetries, or some appropriate
deformations of them,
would persist in highly-curved string backgrounds such as
black holes \cite{wad,witt}.The phase-space-area-preserving
character of the symmetries, then, would guarantee the
maintenance of quantum coherence even during extreme
 physical
processes
like black hole evaporation/decay, for the
reasons
argued in \cite{emn2}.
\pr
To check these considerations explicitly for curved
space-times seems a difficult task. The
weak-field
expansion methods that are used in
extracting the $\beta$-functions of
$\sigma$-model
perturbation theory might not prove sufficient
for getting exact information about
complicated backgrounds
such as black holes.
On the other hand, already
from critical string theory
we know
of cases where exact
conformal field theories
have been used to
circumvent the patterns
of perturbation
theory
around complicated string ground states,
like Calabi-Yau compactified
spaces. Gepner's construction \cite{gep}
of tensoring superconformal field theories
has provided us with non-perturbative information
about {\it exact} solutions of string theory
in the form of Calabi-Yau Ricci-flat spaces,
which
seemed not to correspond
to solutions of perturbative $\beta$-function
equations \cite{pope}.
In Gepner's construction an important method
was the use of Fateev-Zamolodchikov
parafermions
to represent
the pertinent
$N=2$ superconformal algebras. It is in this
representation that {\it extra} selection
rules, not known previously,
have been found for correlation functions in
Calabi-Yau backgrounds \cite{ross}.
\pr
Motivated by these results in critical string theory,
it was
natural to ask
whether
similar constructions of exact conformal field theories
can be found in the case of
string
black hole backgrounds. This question was answered positively
by Witten \cite{witt}, who
has shown that it is possible to describe the
interior of the horizon of a black hole in two-dimensional
space-time using a coset $\frac{SL_k (2,R)}{G}$ Wess-Zumino(WZ)
model, with $G=SO(1,1)$ ($G=U(1)_{compact}$) for Minkowskian
(Euclidean) black holes. The important point is that
this is the first time that a {\it finite} theory, like the WZ model,
is used
to describe a target space-time singularity
\footnote{However, it should be stressed
that
the conformal field theories
in question are not a mere ansatz for describing
known local field theory objects
like traditional
Schwarzschild solutions of Einstein's equations.
On the contrary, they serve
to
demonstrate
the fact
that objects
{\it resembling} the space-time singularity structure
of spherically-symmetric non-rotating
four-dimensional black holes \cite{emn4}
appear
as {\it exact} solutions of subcritical
string theory \cite{aben}. This is
similar in spirit to
Gepner's demonstration
that Calabi-Yau (Ricci-flat) spaces are exact (non-perturbative)
solutions of critical string theory.}.
And it is this kind of singularity
that thwarted
earlier attempts to
describe
black holes, or in general
singular objects surrounded by event horizons, in a way
consistent with quantum mechanics \cite{hawk}. Conformal
field theory provides, as we shall
see in section 3, the
consistent construction of a scattering
matrix to describe scattering of
propagating string states (tachyons)
off the black hole background. This is to
be contrasted with the
situation
in point-like field theories,
where such a scattering
matrix was argued not to exist \cite{hawk,ehns}.
\pr
In parallel to
the parafermion representation of Gepner's spaces,
one
could
expect a similar construction
here, as was indeed
shown
in \cite{dixon}. Any state that admits an expansion in terms
of an $SL(2,R)$ current-algebra basis
can be represented as a highest weight
$N=2$ superconformal state,
which admits a parafermion realisation \cite{lyk}.
In bosonised parafermion language, an $SL(2,R)$
algebra is realised by three bosons. To
describe a coset requires factoring out one of them,
and one is left with a two-boson realisation
of the coset model. In complex notation, let $\phi (z)$,
${\overline \phi (z)}$ be the two bosons, with
$j = \partial _z \phi$, $ {\overline j}= {\overline \partial _z \phi}$
the corresponding currents. The parafermion currents are
expressed as
\be
\psi _{+}={\overline j} e^{\phi + {\overline \phi}},
\psi _{-} = j e^{-(\phi + {\overline \phi} )}
\label{paraf}
\ee

\noindent The $W$-symmetry structure is revealed
by
looking at the classical
OPE
between parafermion currents $\psi _{+}(z)\psi _{-}(w)$,
for $z \rightarrow w$ \cite{kir}.
Expanding in powers of $\epsilon \equiv z-w$
and keeping {\it all} powers of $\epsilon$
\be
   \psi _{+}(z)\psi _{-}(z + \epsilon) =
\sum_{r=0}^{\infty} u_r (z) \frac{\epsilon ^r}{r\!}
\label{wsymm}
\ee

\noindent one discovers a world-sheet $W$-algebra,
generated by the commutation relations of the
$u_r$ \cite{wu,wu2}.
Parenthetically, but importantly,
we notice that the coefficients $u_r$
are world-sheet currents of conformal spin $r+1$ \cite{wu2},
which
appear as expansion coefficients
in the
pseudo-differential
operator
$L=\partial _z +\sum_{r=0}^{\infty} u_r (\partial _ z)^{-r-1}$,
defining
the so-called
$KP$
Hamiltonian
basis. In this sense the appearance
of $W$-symmetries is related to the $KP$ hierarchy,
which might be suggestive of new ways of
approaching $c=1$ string theory \cite{wu2}.
In the coset model, the
symmetry algebra generated by the currents $u_r$
is related to the second Hamiltonian
structure of the
$KP$ hierarchy \cite{wu}.
It is actually
a non-linear deformation \cite{wu2}, ${\hat W _{\infty}}$,
of the centerless $w_{\infty}$ algebra
of Bakas\cite{bak}. We shall come
back to this point in section 4. Quantisation
of the model does not simply
require
{\it normal ordering}
these
OPEs, but a {\it redefinition} of
the currents
\cite{wu} so as to ensure the closure of the
quantum ${\hat W}$-algebra. Details of the
construction can be found in \cite{wu}.
\pr
An infinite set of commuting quantum $W$-charges is
constructed as world-sheet spatial integrals
of the
currents $u_r$ (or rather an appropriate redefinition, $W_r$,
in the notation of \cite{wu2,kir}),
with
integer
conformal spin
$s \ge 2$ \cite{wu}
\be
   [ \int dz W_s(z), \int dw W_{s'}(w)]=0 ; s,s' \ge 2.
\label{charges}
\ee

\noindent The charge $W_2$ coincides with the
Hamiltonian of the model, or,
in the case of closed strings,
with the holomorphic part of the $L_0$ Virasoro operator.
Notice also
that spin-one objects
are not included, by construction \cite{kir,wu},
in this set and so the
deformed algebra is not of $W_{1+\infty}$ type
(which is generated by integer
conformal spins $s \ge 1$). The $W_{\infty}$
algebra is known \cite{wu2} to be a subalgebra
of $W_{1+\infty}$. However, it is the $w_{\infty}$
algebra
that has
a well-known geometrical
interpretation as a
phase-space area preserving
symmetry \cite{bak}, and so for our purposes
\footnote{It should be
mentioned,
though, that in view of the matrix-model
result \cite{ms}, one might expect
the actual
target space-time symmetry of the two-dimensional
string theory to
be bigger, allowing for
symmetries generated by conformal spin-one currents.
This would include the canonical deformations
mentioned above. Of course, one cannot exclude the possibility
that the flat space-time symmetries are
larger
than those of the black hole background.}.
it is
sufficient to concentrate on $w_{\infty}$ and its
quantum deformation ${\hat W} _{\infty} $.
\pr
The above constructions are valid for coset WZ models
whose level parameter $k \ge 2$ \cite{kir,wu}.
The model with $k=\frac{9}{4}$ admits an interpretation
as a {\it critical} string theory propagating in a black
hole background \cite{witt}.
The interesting point is therefore the
lifting of this enormous
world-sheet symmetry to a
physical gauge symmetry of two-dimensional
strings. Due to the higher conformal spins of the
generating
currents, it turns out that the induced deformations
about the pertinent background are not in general
canonical. As argued in \cite{gian} recently,
canonical deformations are only a part of the
enormous infinite set of target gauge symmetries.
The requirement that
a field
be $(1,1)$
implies usually an equation of motion {\it and }
some gauge conditions (constraints), for
higher-spin states.
The most general set of deformations discussed
in \cite{gian} relaxes the requirement of
gauge fixing, and hence one
works in arbitrary gauges, so the only remaining
constraint is
conformal invariance. The main conclusion then is
that
any symmetry on the world-sheet, generated
by a current of arbitrary conformal spin, can be
viewed as generating a symmetry of target space-time,
provided translational invariance is maintained on the
world-sheet. Therefore
the corresponding charge
operator should commute with $L_0 - {\overline L_0}$:
\be
   \{ h \} : [h, L_0 -{\overline L_0}] =0
\label{set}
\ee

\noindent The cost
one pays is the introduction of auxiliary
fields
that are pure gauge artifacts
introduced to count
correctly
degrees of freedom \cite{gian}. In the two-dimensional
case, as we shall discuss
in the next section in some detail, the
existence of discrete modes in higher spin levels,
which are associated with
the target $W$-symmetries, emerges precisely from
a relaxation of constraints and/or gauge conditions
which occurs at particular values of energy and momentum
\cite{pol}. Moreover,
the world-sheet charges (\ref{charges}) do satisfy
(\ref{set}) by construction, and therefore they are
responsible for the generation of string gauge
symmetries in physical space-times. In the context
of first-quantised string theory, these symmetries
will be expressed through complicated redefinitions
of the $\sigma$-model fields \cite{emn2}.
{}From the commutation relations (\ref{charges})
it becomes clear that the induced deformations vanish
when integrated
over the
world-sheet. This
should be intuitively expected. The ordinary
target-space general coordinate transformations
are in fact {\it total} derivative effects on the
world-sheet,
being mainly responsible for the difference of
local from global conformal invariance (c.f. the
difference
of $\beta$-functions from ${\overline \beta}$-functions
is expressed as
a general coordinate transformation \cite{tsey}).
Also,
from
a glance at the commutation relations of the charges
(\ref{charges}) with the string level operator in the
corresponding string theory, it becomes evident
that for many of them there is a non-vanishing
(and rather complicated)
result,
thereby implying a mixing of string levels, as expected
for a stringy gauge symmetry transformation.
\section{Selection rules for scattering
in flat space-time }
\paragraph{}
Here we
review briefly the situation in the
flat space-time $c=1$ string. The latter is
interpreted {\it both}
as the asymptotic (spatial infinity)
form of the black hole space-time as well as
the end-point (temporal infinity)
of
perturbative black hole evaporation \cite{witt,emn3}.
We shall put emphasis
mainly on the physics
of this background rather than the mathematical details.
The important point is that the $c=1$ conformal
matter theory coupled to Liouville gravity
admits a space-time interpretation with
the Liouville field $\phi$
being considered as the
{\it spatial}
coordinate, and the Feygin-Fuchs representation $\chi$
of the matter
field as the temporal one
\footnote{An alternative interpretation of the model
as describing
physics in a Euclideanised
two-dimensional universe,
with the Liouville field
being the Euclidean time,
seems not to be compatible with the interpretation
of the model as an asymptotic form of the black hole
solution, where the Liouville field
clearly plays
the r\^ ole
of the
radial (spatial)
coordinate \cite{witt}.}.
Viewed as a flat space-time string theory,
the $c=1$ Liouville-matter system
resembles that of an ordinary subcritical
string theory with an anomaly $Q=\sqrt{\frac{c-25}{3}}=2\sqrt{2}$
in the
Liouville
sector,
but with
{\it imaginary} ``Liouville energies'' \cite{pol}.
For example, the propagating {\it massless}
modes of the system, called
misleadingly `tachyons', turn out to be
described by $(1,1)$ vertex operators of the form:
\be
V^{T}=e^{ik_\mu X^{\mu}}
\label{tach}
\ee

\noindent where $X^{\mu}=(\phi, \chi)$
and $k^{\mu}=(i\epsilon (p), p)$.
The requirement that the operator $V^T$ be
a $(1,1)$ field implies an {\it equation of motion}
$k^{\mu}(k+Q)_{\mu}=-2 $ (with $Q_{\mu} =(iQ,0)$)
for the tachyon field, which is
the analogue of the `on-shell condition' in ordinary string models.
The solution requires that the Liouville energies
are not independent of the matter momenta p, but rather:
\be
     \epsilon (p) = -\sqrt{2} + \alpha  p
\label{energ}
\ee

\noindent where $\alpha = \pm 1$. Notice that $p$ denotes the
{\it vector} and {\it not}
the {\it magnitude} of the matter momentum.
In view of the space-time
interpretation of the matter momentum as energy,
and the Liouville energy as (radial) momentum, the
two signs in (\ref{energ}) indicate the two possible directions
(outgoing and ingoing) of the radial momentum (this is also true
in the case of
spherically-symmetric four-dimensional space-times). As
remarked in \cite{polch},
in Liouville-gravity only
states with momentum $p_{\phi} \equiv \epsilon \ge -\sqrt{2}$
are kept, since these are the only ones that have well-behaved
asymptotic behaviour in the limit where the Liouville
field, {\it assumed } as a free field, approaches $-\infty$.
Hence in (\ref{energ}) $\alpha p > 0$ for {\it both} signs of $p$,
thereby fixing a sign for $\alpha$ for a
fixed sign of $p$. This
defines uniquely the concept of an `ingoing' or `outgoing' particle
in this picture.
\pr
In addition to the tachyons, which are propagating light particles,
the $c=1$ string theory has an {\it infinity} of discrete
(higher-spin) excited states with definite energy and momentum.
These states make their
appearance in the subleading terms of
OPE of tachyon vertex operators
\cite{gross,sak}.
Their presence is essential for perturbative {\it unitarity}
of the scattering matrix $S$ of the $c=1$ string theory.
The discrete states do not appear as external on-shell states
due to their quasi-topological nature; they are rather
soliton-like
states that appear as exchanged states in amplitudes
of tachyons,
and can be found by factorisation \cite{klepol}.
There appear to be
two branches of discrete states \cite{klepol},
one corresponding to the degeneration of gauge transformations
associated with the given string level, and the other to the
relaxation of the constraints imposed
on the polarisation tensors
of the states for generic momenta. The selection rules
discussed in this section refer to the first branch.
The reason is simple. In flat space-time strings
these states are {\it singular} gauge transformations,
and so can be regarded as physical gauge particle states that
{\it interact} with the {\it massless} (propagating)
degrees of freedom, the tachyons. The other branch
seems to
play an important r\^ ole in highly-curved backgrounds.
For instance, at level $1$
the states of the second branch have momentum $k_\mu =-Q_{\mu}$,
and they constitute the excitations of the limiting (vanishing
mass) black hole solution \cite{wad,emn1}.  So the natural
question concerns the physical independence of these branches
of string states. As we mentioned above,
 from the point of view of free-field
 Liouville gravity,
a state with momentum $p_{\phi} \equiv \epsilon
 <   -\frac{Q}{2} $ is excluded \cite{polch},
due to bad asymptotic behaviour in the limit where the Liouville field
$\phi \rightarrow -\infty$.
The situation of course is not clear
if boundaries are imposed on Liouville space from below,
which appears
necessary due to the r\^ ole of the Liouville field as a
cut-off on the world-sheet \cite{polch2,mav}.
We shall resolve
such issues
in the
context of the next section.
As we shall discuss \cite{distl}, the other branch seems to
correspond to equivalent $SL(2,R)$-isospin representations,
and again the discrete states are the ones that look like
singular gauge transformations.
\pr
Before going into details of the selection rules,
it is essential to
discuss some features
of the tachyon amplitudes in this string model.
The most important (and probably mysterious) of them
is the fact that the only non-vanishing amplitudes
for generic values of the momenta of external
tachyons are the ones with signatures
$+++...+-$ or $---...-+$,
where
the sign is that
of $\alpha$ in (\ref{energ}).
{}From a physical point of view this is very strange,
since it implies that only amplitudes
with
either
one
incoming or one outgoing
particle are non-vanishing for generic
tachyon momenta. The remaining amplitudes
vanish
modulo
contributions that are $\delta$-functions
of the momenta \cite{klepol}. The
explanation
we
offer is
the target space $W$-symmetries.
There exist
Ward identities in target space-time,
which are the continuum analogues of corresponding
identities found in the $c=1$ matrix model \cite{ind}.
We argue that these
are
responsible for
the above
selection rules
for
particle scattering in subcritical
string theory. Given that the $W$-symmetries
have their origin on the world-sheet, these
Ward identities are similar in nature to those found in critical
string theory \cite{ven}, and
correspond
to redefinitions
of the target space coordinates. As we have already mentioned,
a similar in nature
(but technically
more complicated )
redefinition of the $\sigma$-model
fields operates here as well \cite{emn2}.  These
issues certainly
deserve more rigorous
treatments than our
simple arguments,
and we hope to come back to them in a future
publication \cite{emnpr}.
\pr
In addition to these strange selection rules,
there are
additional ones
for resonant amplitudes of the form $+++...+-$ (or its mirror),
involving resonances of discrete states, which
{\it restrict} the number of light particles interacting
with them. For
instance, consider the $N$-tachyon amplitude \cite{pol}
\be
  A_{++++...+-}(p_1,...,p_{N-1})=\frac{1}{(N-3)!}
\frac{\Gamma (1-2p_1)}{\Gamma (2p_1)} ...
\frac{\Gamma (1-2p_{N-1})}{\Gamma (2p_{N-1})}
\label{reson}
\ee

\noindent It is worth
pointing out that this is
only the `resonant' part of the amplitude as far as
Liouville energies are concerned. In Liouville
theory coupled to matter, due to the
peculiar Liouville dressing of matter operators,
there is not in general conservation of
Liouville energies. However,
the amplitudes have
`resonant' forms \cite{pol},
for the description of which
free-field methods are applicable. This means that the
boundaries in Liouville field $\phi$-space
are ignored and the Liouville field is treated as a free boson.
All the physical amplitudes are {\it assumed} to be the residues
of Liouville energy conservation poles, and hence
{\it both} energies and momenta are assumed conserved, as in
ordinary string theory, implying $SL(2,R)$ invariance in the
usual Koba-Nielsen form. The
kinematics then implies:

\bea
\nonumber \sum_{i=1}^{N-1} p_i &=& \frac{N-2}{\sqrt{2}}  \\
p_N &=&-\frac{N-2}{\sqrt{2}}
\label{cons}
\eea

\noindent where one should be careful to include
the anomaly $Q$
in the Liouville energy conservation law.
Combinining then energy and momentum conservation
one finds that one of the momenta is completely
determined, for instance $p_N$.
\pr
The amplitude (\ref{reson}) has `leg' poles at momenta:
\be
         p=\frac{n}{2} ; \{ n \}=Z^{+}
\label{poles}
\ee

\noindent In usual string theory,
poles in external
particle states are explained by factorising the
amplitude using the OPEs
of two vertex
operators approaching each other on the world-sheet.
A similar thing happens here \cite{sak}. When
a pole occurs, say in the $(1,N)$ particle-channel, then
the conservation equations imply restrictions
on the number $N$  for a given exchanged discrete
state.
\pr
The discussion is facilitated if one uses a more
convenient set of quantum numbers to classify
the discrete states, rather
than merely their spin and
string level number
as
in ordinary (critical)
string theories. Due to the low dimensionality
of the target space-time,
it turns out that
the string states can be
classified
by assembling them in $SU(2)$ multiplets \cite{klepol}.
At this point,
it is worth
making a
digression
which will be useful for
gaining some intuition about
the classification of states around more complicated
backgrounds like black holes, which will be discussed
in the next section.
\pr
The value of the central charge of the conformal field theory
plays a crucial r\^ ole in the sort of representation
one uses to represent the string states. Fateev and
Zamolodchikov have shown that for $c < 3$ unitary CFT
one could convert the parafermion algebra used to describe
the states into a current algebra of $SU(2)$.
On the other hand, for
$c > 3$ one has to use $SL(2,R)$ non-compact current
algebras \cite{dixon}. We shall come back to these issues in section 4.
For our purposes, the states of the $c=1$ flat space-time
string theory can be assembled in $SU(2)$
representations, characterised by two
quantum numbers, the SU(2) isospin $j=\frac{n}{2}$, $n$ integer,
and the magnetic number (its third component) $m=-j,-j+1,...,j$.
Following Klebanov and Polyakov \cite{klepol},
and using their normalisation in which the two-momentum
is rescaled by a factor of $\sqrt{2}$,
the conformal invariance condition for an operator
corresponding to a discrete state after Liouville
dressing $e^{\epsilon (j) \phi}$ reads
in terms of $j$,$m$ as:
\be
          j^2 - \epsilon (\epsilon + 2) =1-N
\label{su}
\ee

\noindent for a level $N$ state. From this and the analogy
with string theory,
where the left-hand
part of (\ref{su}) is
just the two-momentum squared,
one finds the analogy
between the two pictures:
\bea
 \nonumber     p \rightarrow m   \\
               \epsilon ^{\pm} \rightarrow -1  \pm j
\label{anal}
\eea

\noindent The two signs correspond to two different
branches of discrete states, corresponding either
to singular gauge transformations or to a relaxation
of constraints imposed (for generic momenta) on the
polarisation tensors of higher excited states.
The discrete states that lead to selection rules
in flat space-times are the ones that correspond to
singular gauge transformations, which are those with
the positive branch in (\ref{anal}).
For more details we refer the reader to the literature
\cite{klepol}.
\pr
The selection rules on amplitudes (\ref{reson})
follow immediately
upon factorisation and use of
(\ref{poles}). At each trilinear OPE vertex
there is
exact energy-momentum conservation
without the screening charge $Q$. This
is
attributed as a $Q$-insertion to the remaining part of the
diagram. Using the analogy (\ref{anal})
we get $2p_1 = j + m +1 =2j - (N-4)$. From (\ref{poles})
we get the following selection rules for the emission
of the $(j,m)$ state \cite{klepol}:
\bea
\nonumber j &=& m + N - 3  \\
                        j &\ge &  \frac{N-3}{2}
\label{sel}
\eea

\noindent where $N-2$ is the number of particles
{\it interacting} with the discrete state $(j,m)$.
{}From (\ref{sel}) we observe that in
a scattering
resonant
process where an intermediate topological
state $(j,m)$ is excited, the number of emitted particles is
restricted, or vice versa,
in a process where $N$ particles are scattered,
the quantum numbers of the internal states are
restricted
according  to (\ref{sel}). In
the space-time interpretation
of this theory as a spatially-asymptotic form of a
black hole, this
kind of selection rule
yields
some information about the target-space gauge symmetries
characterising the black hole configurations ($W$-hair).
In view of the group-theoretic
equivalence of the two branches of discrete
states in that model, the corresponding selection rules
can be
considered
physically complete.
We shall return to these issues in the  next section.
At present,
we remind the reader
that the states of the
$c=1$ string
form a $W$-algebra that admits a phase-space
area-preserving interpretation \cite{klepol,witt2}.
The selection
rules given above, both the vanishing of $+++...--...-$ amplitudes
as well as the rules $(\ref{sel})$,
are a
manifestation
of this symmetry. In purely flat space-time backgrounds,
of course,
the resulting charges are {\it trivial}. It is
when one considers more complicated string vacua, as we shall
do next, that the corresponding selection rules give
information about the
gauge charges that
characterise, say, the black hole. In that sense, the rules
(\ref{sel}) are part of
the asymptotic form of selection rules for
scattering light particles off the black hole background
by
excitation of black hole
topological (internal)
degrees
of freedom.

\section{Selection rules for
scattering off
black holes}
\paragraph{}
The above
situation changes significantly when
we describe
highly-curved string backgrounds
such as
black holes. Witten showed \cite{witt}
that there is an exact description of such objects
as
coset WZ models
for a given value,
$k=\frac{9}{4}$, of the level parameter. In contrast
to
Liouville gravity, for this value of $k$ the
conformal field theory {\it per se} has central charge 26
to compensate the ghost
contributions.
Therefore,
in view of the Fateev-Zamolodchikov
result \cite{dixon} one expects the compact $SU(2)$ picture
for the classification of states to break down. Indeed,
as discussed in \cite{dixon},
the appropriate classification
can be made via $SL(2,R)$ isospin representations, as we mentioned
in section 2. The contact with Liouville gravity, discussed
in the
previous section,
is made only {\it asymptotically}, either at spatial infinity
or in the limiting (zero-mass) black hole case \cite{emn1,wad}.
However, it should be stressed that the coset string model
seems to be characterised by more states than the flat $c=1$
Liouville model. A rather complete discussion of the
spectrum is given in \cite{distl} where we refer the reader
for mathematical details. Here we shall
concentrate on
some aspects
of that work relevant for obtaining
selection rules revealing the underlying
$W$-symmetry structure of the target
space \cite{emn2,kir,wu}. We should
emphasize that there are differences \cite{distl,emn3}
between Minkowskian and Euclidean black holes, viewed
as coset conformal models. One is {\it not}
obtained from the other
via a
simple Wick rotation. In particular,
the spectrum of physical states is different. There are more
states in the Euclidean case, which is natural from the point
of view of the
target space-time interpretation, due to
the compactification of the (target)
`time' coordinate.
Some aspects do transcribe simply, though,
from Euclidean to Minkowski space, and it is in this
sense that
it is useful sometimes
to formulate the
problem in
Euclidean space,
where it is easier mathematically, and then
pass to
the Lorentzian signature formalism.
Here we shall follow the latter method;
the simple Wick rotation
is
sufficient for the selection rules.
\pr
Following Distler and Nelson \cite{distl},
we characterise the states in the coset model
$\frac{SL_{\frac{9}{4}}(2,R)}{U(1)}$ by two quantum numbers,
the $SL(2,R)$ isospin $j$
and its third component $m$, the latter being defined
by
the
diagonalisation of the generator of the
$U(1)$ current algebra.
The
requirement for a state to be a Kac-Moody
primary leads to the `on-shell' conditions \cite{distl}
\be
(L_0 - 1)|j,m> = (-4j(j+1) + (\frac{2m}{3})^2 -1 ) |j,m> =0
\label{onshell}
\ee

\noindent which can be solved to yield
\be
       j= -\frac{1}{2} \pm \frac{1}{3}m
\label{root}
\ee

\noindent Higher-level string states, corresponding to
non-zero
`oscillator number' ${\it N}$, also exist, but at discrete
values of $j$ and $m$. We shall come back to them later on.
The lowest oscillator number states
are just the
`tachyons' of the
two-dimensional string theory.
{}From the asymptotic connection with
flat-space two-dimensional strings one can make the
following {\it analogy} \footnote{It should be stressed
that this is only an analogy, or an approximation at spatial
infinity perhaps. The states of the coset model cannot be
represented as {\it plane waves} in target space,
due to the highly-curved space-time structures.}
between the quantum numbers $(j,m)$ and the
plane wave two-momentum $(i\epsilon_\phi (p_x), p_x)$
\bea
 \nonumber
     j    &  \rightarrow & \frac{\epsilon_{\phi}}{2\sqrt{2}}  \\
     m  &\rightarrow & \frac{3 p_x}{2\sqrt{2}}
\label{plane}
\eea

\noindent where $p_x$ denotes the momentum corresponding to
the matter part of the conventional $c=1$
Liouville-matter system. In
the space-time black hole
interpretation of the
model \cite{witt}, the
Liouville field would play the r\^ ole of (spatial)
radial momentum and $p_x$ the r\^ ole of energy.
The two signs therefore in (\ref{root}) indicate
the outgoing or ingoing direction of the (asymptotically plane)
wave front.
\pr
It is important to notice that
$SL(2,R)$ representations
are classified  by the Casimir $j(j+1)$ rather then $j$ alone, so
the physically {\it inequivalent} representations are
restricted (for the
Euclidean model where $j$ is real) to
\be
     j \ge -\frac{1}{2}
\label{ineq}
\ee

\noindent From the analogy (\ref{plane}) we then recover the
restriction
$\epsilon_\phi \ge -\sqrt{2}$,
familiar from $c=1$
Liouville gravity as
discussed in section 3.
It is this group-theoretic
equivalence of representations that
leads to the conclusion that the
 only physically inequivalent states are those corresponding to
singular gauge transformations. This is consistent with the
asymptotic connection of the model to
the flat $c=1$ string
theory, as well as the fact that it is this branch of
discrete states that seems to be responsible for the
$W$-symmetry structure \cite{klepol,emn1,wu,kir}
which is {\it common} to both theories.
\pr
{}From the asymptotic Liouville-like intepretation
it is also clear that similar selection rules to
those applying in the previous section
also apply here. For instance, the
vanishing, for generic momenta, of tachyon
amplitudes of signatures $+++...--...-$
(or $---...++...+$) also occurs here, and can be
thought of, along with the additional selection rules
referring to amplitudes with $++...-$ (or $-++...+$) to
be discussed below, as a manifestation of the $W$-symmetry
characterising the target space of the coset model.
At this point it should be mentioned that
the very fact that this symmetry is of {\it similar nature}
as that of the $c=1$ Liouville-matter system is an additional
indication of an underlying `hidden' gauge symmetry structure
that describes evaporation/decay of the black hole space-time
and its evolution to reach asymptotically (in time) flat target
spaces. We shall briefly comment
on this important issue in
section 6.
\pr
The selection rules for the non-vanishing amplitudes
$+++...-$ (or its mirror) are obtained in a similar way
as those of section 3. Some remarks
are in order, however.
The first
concerns
`energy-momentum' conservation in this
context, which
would again imply that
from the point of view of Liouville-theory the
correlation functions of the coset model would reproduce
only the resonant parts of the amplitudes,
where free-field
methods are applicable.  That this
is
indeed the case becomes
evident  from the
free-boson realisation of the coset model \cite{kir}.
The generic vertex operator $V^{j}_{m}$
of the (Euclidean)
coset theory, corresponding to a state
characterised by
the two quantum numbers $(j,m)$,
is represented as
\be
     V^{j}_{m}=e^{\frac{8im}{9}\chi} Z^{j}_{m}
\label{coset}
\ee

\noindent where $Z^{j}_{m}$ is a primary of the $N=2$
superconformal theory used to represent the $SL(2,R)$ current
algebra \cite{dixon}, and $\chi$
is a free boson. It is then clear
that in correlation functions for
operators of the form
(\ref{coset}) there is
conservation of the $U(1)$-charge
$\propto$ $m$, as a result of the boson zero-mode integration.
In view of the on-shell condition (\ref{root}),
this also
implies $j$-isospin conservation. The plane wave analogy
is then transcribed
into energy-momentum conservation. It is important,
however, to notice
that the space-time interpretation of the model requires
the presence of a Liouville screening operator $\epsilon_\phi
= -\sqrt{2}$ in correlation
functions. This means that a $Q$-anomaly insertion term
will always be present in Liouville energy ($\epsilon_\phi$)
conservation
laws.
\pr
We now briefly review the spectrum of the Euclidean coset theory
found by Distler and Nelson \cite{distl} ($r,s$ are integers):
\bea
\nonumber {\tilde D}^{\pm} -  series   &:&
(\epsilon_\phi , p_x)
=(\frac{1}{2\sqrt{2}} (2s+4r-5), \pm \frac{2s-4r-1}{2\sqrt{2}}) \\
\nonumber  D^{\mp}-  series   &:&
(\epsilon_\phi , p_x)=
(\frac{1}{\sqrt{2}}(s+2r-3) , \pm \frac{s-2r+1}{\sqrt{2}}) \\
     C-  series   &:& (\epsilon_\phi , p_x)=
     ( \sqrt{2}(s+r-1) , \frac{2(s-r)}{\sqrt{2}})
\label{spec}
\eea

\noindent By appropriate definition of the inner product of
states \cite{distl},
one can pass to
the Minkowskian
formalism by analytic continuation.  Not
all the discrete
states
found in Euclidean black holes
can be
mapped into the Minkowski case. This is true {\it only} for the
states
with $m = \frac{n}{2}$,
where
$n$ is an
integer.
As
is clear from (\ref{spec}),
there are {\it extra}
states in the coset model as compared to the Liouville theory.
For instance,
part of the ${\tilde D}^{\pm}$
corresponding to
$\frac{3}{8}(2s-4r-1)=\frac{n}{2}$, with $n$ integer,
exist in the Minkowskian black hole but not in the
corresponding $c=1$ flat string theory. The
screening operator ($r=s=0$)
exists only in the series $C$,
which contains an integer subset
of the
discrete
values of tachyon momenta that lead to poles in amplitudes
\cite{pol}. In the
context of the
scattering of light particles off the Euclidean
black holes,
these discrete tachyonic values are interpreted
\cite{verl} as bound states of the string tachyon mode
with the black hole background. In the context of Minkowskian
formalism
such states are completely {\it absorbed} by the
black hole.
This can
easily be
seen
by looking at the reflection coefficient $R(p)$
of the
quantum-mechanical
scattering process. It assumes the form \cite{verl}
\be
   R(p) = \frac{\Gamma (1-ip)\Gamma ^2 (\frac{1}{2}+2ip)}
   {\Gamma (1+ip) \Gamma ^2 (\frac{1}{2} + ip)}
\label{refl}
\ee

\noindent For $p=i\frac{n}{2}$, where $n$ is a {\it positive}
integer, the $R$-coefficient vanishes. The fact that these values
are purely imaginary
is a consequence of
the Lorentzian
signature. In the
Euclidean black hole case
the corresponding
momenta are
positive integers or half integers.
{}From a
purely conformal-field-theoretic point of view
one can consider now the OPE of vertex operators corresponding
to these discrete values. Asymptotically in space-time
this would imply factorisation of the scattering matrix
to one of the discrete states. This interpretation can be
carried out in the fully-curved black hole background
if one assumes that, in an
experiment scattering a
massless particle off the black hole background,
the absorbed light particle
excites a given discrete massive
mode of the black hole.
The latter then decays by emitting particles whose number
is restricted by the nature of the excited massive states.
To see this one can repeat the steps of section 3,
assuming that the momentum of the
light particle takes
one of the discrete values
that lead to {\it bound} states.
The net result for the selection rules is
\bea
\nonumber     j=\frac{1}{3}m - 1 + \frac{1}{2}(N-2)  \\
        j \ge \frac{1}{4} (N-5)
\label{selbh}
\eea

\noindent Notice the correspondence with the
flat space selection rules (\ref{sel}) upon making
the analogy
\bea
\nonumber     2j_{SL(2,R)} \rightarrow j_{SU(2)}-1  \\
              \frac{2}{3}m_{SL(2,R)} \rightarrow m_{SU(2)}
\label{sutwo}
\eea

\noindent However (\ref{sutwo}) should only be considered
as a formal correspondence. The set of discrete states
in coset theory is larger than that of flat $c=1$ strings,
and hence there are always {\it extra} selection rules in
the former model.
\pr
The above selection rules have been originally
derived for the case of the Euclidean black hole .
To pass into Minkowski formalism one notices the following.
{}From the cohomological approach of Distler and Nelson \cite{distl}
it becomes clear that the higher-level discrete states
in this string theory are characterised by {\it real j}.
The usual passage from Euclidean to Lorentzian black holes
is then given by the replacement $m \rightarrow i\mu$,
where $\mu$ is defined by the diagonalisation of the generator
of the $SO(1,1)$ subalgebra in the coset model
$\frac{SL(2,R)}{SO(1,1)}$.
For the discrete states $\mu$ is taken to be purely imaginary,
$\mu=-im$, so both  selection rules (\ref{selbh}) can be
translated directly into the Minkowskian theory.
Again, however, in view of the {\it inequivalent} spectra
of the two theories the physical
content of the (formally identical)
selection rules changes.
\pr
The physical interpretation of these selection rules is now clear.
They
constitute a manifestation of the
underlying $W$-symmetry structure of the theory.
The
non-trivial conserved charges
that characterise the background (black hole)
play an analogous r\^ ole to {\it topological }
charges in conventional local field theories, labelling
not particles but rather vacuum sectors of the theory
where
the scattering of particles takes place. The key r\^ ole
of the
$s$-wave dynamics and the selection rules are
reminiscent of the Callan-Rubakov \cite{caru}
process in scattering off a monopole. Further we
recall that
the instanton
number
in the
standard electroweak model
also leads to selection rules
restricting the number as well as the
flavour
of emitted particles, for
given
incoming states \cite{ringw}. In our case, since
there is only
one flavour of particle,
the only
restriction concerns the number with which the exchanged state
can interact. In contrast to the local field theory
example, however, here we have an {\it infinite} set
of selection rules that differ for each internal state, and an
{\it infinity}
of topological (conserved) charges which label
the internal degrees of freedom of the black hole.
It is this feature
that explains naturally the large statistical
entropy of the latter, with the entropy
being defined \cite{emn4}
as the loss of information for an observer at spatial
infinity  who, ignoring the higher excited black hole states,
measures only the classical energy and charge of a non-rotating
black hole. However, these can in principle be distinguished,
for example, by measuring the number of particles the black holes
emit when they decay. We discuss another class of distinguishing
measurements in the next section.

\section{Aharonov-Bohm measurements}
\paragraph{}
We have argued previously that the discrete
states appearing both in the flat space $c=1$ string
and in the black hole coset model are {\it singular}
gauge transformations (or physically equivalent
representations thereof). This picture is consistent
with that stemming from ordinary strings,
where the various states are viewed as gauge particles
\cite{grosmend}. In higher-dimensional string theories,
the fact that
stringy symmetries {\it mix} the various mass levels implies
{\it spontaneous} breaking of these symmetries.
In two dimensions spontaneous breaking cannot occur
\cite{col},
so the gauge $W$-symmetries
constitute a set of {\it unbroken} string gauge symmetries.
The selection rules discussed above are a
manifestation
of this. In topologically non-trivial backgrounds, like the
Schwarzschild black hole, these gauge symmetries
lead to an infinity of
{\it non-trivial} conserved charges. The association
of the latter with discrete (non-propagating) delocalised
string states implies their {\it topological} nature.
Formally this can be seen as follows. At
the level of the target-space effective action
the coupling of the (gauge) string states with
the conserved currents $J^{M...}$
that characterise the string background
can be represented generically as
\be
      \int _{space-time} \sum_{s=string-states}
      A_{M...}^{s} J^{M...}_{s}
\label{coupl}
\ee

\noindent In the
spherically-symmetric case the integral over
the four-dimensional space-time reduces to a two-dimensional
integral. In that case,
the string states $A$ are singular gauge
transforms $d\Lambda$,
and the currents $J$ are induced by
the world-sheet currents leading to the charges (\ref{charges}).
{}From
current
conservation it follows immediately that
the terms (\ref{coupl}) are purely {\it surface} terms,
and therefore they can only carry {\it global} (topological)
information about the space-time manifold.  This is consistent
with the fact that these charges are {\it exactly} conserved
in highly-curved space-times. In fact it is the only possibility.
\pr
The pure gauge nature of the string states that couple to the
conserved quantum numbers of the black hole leads to a better
analogy with ordinary Aharonov-Bohm  experiments used
to measure
conventional discrete gauge hair in local field theories \cite{lah}.
Take, for instance, the antisymmetric tensor hair in string-inspired
black-hole theories
involving gauge fields that couple to the
antisymmetric tensor field strength. There
{\it both} the gauge field strength
and the antisymmetric tensor field strength are
identically
{\it zero} outside the horizon. Also the charge
corresponding to the antisymmetric
tensor field strength is a
surface term \cite{lah} $\int _{S(V)}B$, over
a two-dimensional
surface $S(V)$ surrounding a three-dimensional spatial
volume $V$.
A similar thing happens here. The field strengths
corresponding to the string states vanish outside the
horizon, since the states are pure gauge transforms,
with singularities at the origin of the black hole.
The corresponding charges, characterising the black hole solutions,
are also surface terms. Due to spherical symmetry, the
charges are just conserved numbers, since surface terms
in effectively two-dimensional spaces consist just of points.
The surface character of the charges can be easily exhibited
for the first few of them (e.g. the total energy of the black hole
\cite{witt,emn1}).
That these charges characterise
the spatially-asymptotic form of the coset model is another
manifestation of this fact.
\pr
Having established these properties of $W$-hair,
we now
describe
a
{\it gedanken} experiment
for
measuring it. The process will be
similar to that of
antisymmetric tensor hair \cite{lah},
but
with some crucial
differences. Consider the scattering of four-dimensional
fundamental strings off a spherically-symmetric
black hole background.
One considers an interference experiment at a final
point $B$
between fundamental strings that have
travelled from an initial point $A$ via two paths: in
flat space,
and
in a black
hole environment.
The world-sheet of the latter
fundamental string
encloses
the black hole singularity at a
point $C$, and
this produces a surface-like
coupling with
the
correponding charges of the black hole of the form (\ref{coupl}).
The latter being topological
will only manifest itself as an Aharonov-Bohm phase in the
wave function of the fundamental string , which can be measured
via an
appropriate interference experiment at the point $B$.
Due to the fact that the charges in the black hole background are
in correspondence \cite{ms} with the higher-spin states, in
order to measure individual charges one has to use
{\it polarised} fundamental strings where the particular
string mode is excited. Here comes an important
difference from ordinary strings. If we were living strictly
in two dimensions,
these modes could not propagate,
due to their
topological nature, so scattering
these modes
off the
black hole
is not possible in the ordinary sense.
However,
in four-dimensional topologically non-trivial
spherically-symmetric space-times
(whose singularity structure is described by two-dimensional
strings \cite{emn4})
one can consider propagating higher-level string states,
whose $s$-wave sector resembles that of two-dimensional strings,
symmetrywise. In this sense one considers a fundamental
string whose world-sheet
encloses
a black hole at $C$. The fundamental string
is polarised so that
a
level $N$ state is excited, say. The $s$-wave sector of these
states (which is topological) couples to the conserved charges of the
spherically-symmetric black hole and makes a non-trivial phase
contribution to the wave function of the propagating
four-dimensional string at $B$.
\pr
A final comment concerns the formal character
of the process. For the
string propagating in a flat
space-time background         .
one can
use the $c=1$ string theory to describe its
propagation. However, in this
case there will be a
mismatch when one couples the two theories at $C$,
due to the extra states
that the coset model (black hole background) appears to have
\cite{distl}. Hence the most correct treatment, consistent
from a string field theory point of view, would be to treat
the flat space-time as an {\it asymptotic region} of the
black hole space-time. However,
in view of the similarity
of the target space symmetries between the two models, it
is
possible that the non-trivial
conserved charges are in {\it correspondence} with the
set of states of the asymptotic
$c=1$ Liouville matter system. The existence of stringy symmetries
in the coset model {\it mixing} levels
and {\it mapping}, for
instance, the
discrete states ${\tilde D}^{\pm} \rightarrow D^{\mp} $
\cite{distl} offers
support to these arguments, and
probably implies
that the
actual (physical) spectrum of the coset model might be smaller
than it appears. We hope to return to these important
issues in a
forthcoming publication \cite{emnpr}.

\section{Conclusions and prospects}
\paragraph{}
We have argued \cite{emn1,emn2}
that the infinite set of of gauge
symmetries possessed by string theory maintains
quantum coherence in black hole physics, which is impossible
with the finite set
of gauge symmetries present in any local field theory. Specifically,
in the
case of spherically-symmetric four-dimensional string black holes,
which can be modelled using an effective
two-dimensional
string theory, we have argued that there are an inifinite set
of stringy conserved charges, $W$-hair, associated with a
phase-space-conserving $W$-symmetry, that maintain quantum coherence.
In particular, we have shown
that string black holes
decay in the same quantum-mechanical way as conventional
massive string states and that a thermal or mixed description
is not necessary \cite{emn3}. In
this paper, we have shown how $W$-symmetry
enforces selection rules for particle production in the
scattering of a light particle off a black hole,
reminiscent of particle scattering off a skyrmion or a
monopole,
and made the
point that all the $W$-charges (which are topological in nature)
can in principle be measured by Aharonov-Bohm phase interference
experiments for suitable selected massive
flat-space string modes. Thus $W$-hair is in principle
{\it measurable}.
\pr
Several more-or-less technical issues, some very important
and non-trivial, remain to be clarified. These include
a complete characterisation of the symmetries of string theory
in target space-time, and an explicit space-time representation
of the $W$-symmetries which we have argued to be so important.
These $W$-symmetries arise both in the black hole coset WZ
model and in the $c=1$ flat-space model that corresponds to it
at radial (spatial) and temporal (in perturbation theory) infinity.
However, there are unresolved questions about the size of the
coset model spectrum and its relation to that of the $c=1$
model. Moreover, we should emphasize that although
we believe that the $s$-wave part of the full
four-dimensional problem captures the essence
and is now solved, and we have applied our arguments
also to axi-symmetric four-dimensional black
holes \cite{emn4}, we still
attach great importance to proving that quantum coherence
is maintained for generic four-dimensional spaces with
event horizons and singularities.
\pr
However, we attach even greater importance to a more fundamental
limitation of our analysis, namely that it applies in a
given black hole background, and can be extended to discuss
the evolution of a black hole towards the flat-space
$c=1$ model that describes its end-point in perturbation theory,
but we cannot yet describe non-perturbative transitions
between string vacua, and hence we
cannot yet be sure that they also
respect quantum coherence.
To establish the
maintenance of quantum coherence also during tunnelling
processes would require
the extension of the infinite set of stringy gauge
symmetries off-shell. We are optimistic
that this can be done, since
there is no apparent
problem for quantum mechanics in the
$c=1$ string field theory of
Das and Jevicki \cite{das}, nor
for the $c=1$ matrix model that is fully
non-perturbative. However, a more complete understanding
of non-perturbative phenomena in string theory is necessary for the
discussion of relevant issues in black hole physics.
We know that in perturbation  theory a string black hole loses
mass-energy until it resembles a discrete state in $c=1$ model,
but then what ? The fate of a black hole is essentially
non-perturbative.
\newpage
\noindent {\Large{\bf Acknowledgements}} \\
\par
One of us (J.E.) thanks the Cosmic Ray Group of the
Physics Department of the Universidade Estadual de Campinas
for its hospitality during
part of
this
work. N.E.M. would like to thank HARC (Houston) and
the Theory Group of
Los Alamos National Lab.
for hospitality during the initial stages of this work,
and
D.V.N. would like to thank  the Physics Group of the
Rockefeller University
for warm hospitality and discussions  while this work
was in preparation.
The work of D.V.N. is partially supported by DOE grant
DE-FG05-91-ER-40633.
\pr

\end{document}